\documentclass[aps,prb,superscriptaddress,showpacs,twocolumn,amssymb]{revtex4}
\usepackage{graphicx}
\begin{document}

\title{Time-Dependent Density-Functional Theory for the Stopping
Power\\
of an Interacting Electron Gas for Slow Ions}

\author{V.~U.~Nazarov}

\affiliation {Department of Physics and
Institute for Condensed Matter Theory,
Chonnam National University,
Gwangju 500-757, Korea}

\affiliation {Institute for Solid State Physics, University of
Tokyo, Kashiwa, Chiba 277-8581, Japan}

\author{J.~M.~Pitarke}

\affiliation {Materia Kondentsatuaren Fisika Saila, Zientzi
Fakultatea, Euskal
Herriko Unibertsitatea, 644 Posta Kutxatila, E-48080 Bilbo, Basque
Country, Spain}

\affiliation {Donostia International Physics Center and Centro
Mixto CSIC-UPV/EHU, Donostia, Basque Country, Spain}

\author{C.~S.~Kim}

\affiliation {Department of Physics and
Institute for Condensed Matter Theory,
Chonnam National University,
Gwangju 500-757, Korea}

\author{Y.~Takada}
\affiliation {Institute for Solid State Physics, University of
Tokyo, Kashiwa, Chiba 277-8581, Japan}

\date\today

\begin{abstract}
Based on the time-dependent density-functional theory, we have derived
a rigorous formula for the stopping power of an {\it interacting}
electron gas for ions in the limit of low projectile velocities.
If dynamical correlation between electrons is not taken into account,
this formula recovers the corresponding stopping power of {\it
noninteracting} electrons in an effective Kohn-Sham potential. The
correlation effect, specifically the excitonic one in electron-hole
pair excitations, however, is found to considerably enhance the
stopping power for intermediately charged ions, bringing our theory
into good agreement with experiment.
\end{abstract}

\maketitle

The concept of electron correlation permeates many subfields of
condensed matter physics, but it is rather difficult to find good
examples of its clear manifestation. It is even more difficult to
unambiguously display its importance in time-dependent processes.
Here we report a success in obtaining a clean theoretical formula
directly relating dynamical correlation, as characterized by the
exchange-correlation (xc) kernel of time-dependent density-functional
theory (TDDFT), with the experimentally observed stopping power (SP)
of solids for slow ions.

The nonperturbative theory of scattering of {\it noninteracting}
electrons by a central potential combined with the density-functional
theory (DFT) of Hohenberg, Kohn, and Sham~\cite{Hohenberg-64,Kohn-65}
has proved successful in calculating the SP of an electron gas (EG)
for slow ions.\cite{Echenique-81} This scheme accurately reproduces
the measured energy loss of slow protons and antiprotons in a variety
of solids and successfully accounts for the observed oscillatory
behavior in the energy loss of slow ions with increasing charge,
which is known to originate from the formation of closed shells of
bound states around the ion.\cite{Echenique-86} These calculations
were extended to projectile velocities approaching the Fermi velocity
of the target,\cite{Salin-99} but few attempts have been made to
include the effect of many-body dynamical correlation that is absent
in the current DFT schemes.\cite{Nagy2-89}

In this Communication, we address the effect of dynamical correlation
and we employ TDDFT \cite{Runge-84} to derive a rigorous formula for
the friction coefficient (SP divided by the projectile velocity at
its zero value) of an {\it interacting} EG. This formula fully includes
the effect of dynamical correlation through the imaginary part of a
frequency-dependent xc kernel, and it reduces in the absence of
dynamical correlation to the SP of noninteracting Kohn-Sham (KS) electrons
first reported by Echenique {\it et al.}\cite{Echenique-81}

By adoption of the widely used local-density approximation (LDA), our
formula allows us to make a reasonably accurate evaluation of the
friction coefficient of an interacting EG as a function of the
projectile charge $Z_1e$ for $Z_1$ less than about 15.
Our calculations indicate that dynamical correlation considerably
enhances the SP of slow ions, which is due to the excitonic effect
in the electron-hole (e-h) pair excitations. In particular, this
effect makes the ratio of maximum to minimum values in the
$Z_1$-oscillatory behavior of the SP to decrease, leading to better
agreement with experiment.

We consider a recoiless probe particle of charge $Z_1e$ moving with
velocity ${\bf v}$ in an otherwise homogeneous gas of interacting
electrons at zero temperature. The average energy lost per unit
length traveled by the probe particle (the so-called stopping power
of the target) is simply the retarding force that the polarization
charge distribution in the vicinity of the projectile exerts on the
projectile itself. Accordingly, one can write
(we use atomic units)\cite{Echenique-90}
\begin{equation}\label{0}
-\frac{dE}{dx}
=\frac{Z_1}{v}\!\int \! d{\bf r}\,
d{\bf r}'\,\delta({\bf r}\!-\!{\bf v}t)\,
{\bf v}\!\cdot\!{\bf\nabla}_{\bf r}\,n^{ind}({\bf r}',t)
/|{\bf r}\!-\!{\bf r}'|,
\end{equation}
$n^{ind}$ being the electron density induced by the projectile.

Our starting point is the following rigorous expression for the friction coefficient
of an {\em interacting} electron gas:
\begin{equation}\label{stopp}
Q \!
= \! \! \frac{1}{8 \pi^3} \! \! \int \! \! \! d{\bf q} \, d{\bf k} \, v_{\bf q} v_{\bf k}
\frac{({\bf q}\cdot{\bf v})}{v} \! \! \frac{({\bf k}\cdot{\bf v})}{v}
\frac{\partial}{\partial\omega} {\rm Im}\chi_{{\bf q},{\bf k}}(\omega)
\! \Bigr |_{\omega=0},
\end{equation}
where $v_{\bf q}= -4\pi Z_1/q^2$ is the Fourier transform of the
bare Coulomb interaction and $\chi_{{\bf q},{\bf k}}(\omega)$
is the double Fourier transform of the linear density-response function
of a system of {\it interacting} electrons in the static field of
an impurity of charge $Z_1$ at ${\bf r}=0$.
In order to prove Eq.~(\ref{stopp}), we
first expand the right-hand side of
Eq.~(\ref{0}) in powers of $Z_1$ and we then exploit the symmetry properties
of many-fold density-response functions to demonstrate that
this expansion divided by the projectile velocity at its zero value
exactly coincides (to all orders in $Z_1$) with the corresponding expansion
of the friction coefficient of Eq.~(\ref{stopp}).
\cite{Nazarov0}

The density-response function $\chi_{{\bf q},{\bf k}}(\omega)$ depends
on the {\it unknown} ground and excited states of a many-electron system
and is, therefore, difficult to calculate. Nevertheless, TDDFT allows
us to express $\chi_{{\bf q},{\bf k}}(\omega)$ as a solution of the
integral equation\cite{Petersilka-96}
\begin{equation}\label{eq}
\chi_{{\bf q},{\bf k}}(\omega)=\int d{\bf
p}\,\tilde{\epsilon}^{-1}_{{\bf
q},{\bf p}}(\omega)\,\chi^0_{{\bf p},{\bf k}}(\omega),
\end{equation}
where $\tilde{\epsilon}^{-1}_{{\bf q},{\bf k}}(\omega)$ represents
the so-called inverse test-charge--electron dielectric function,
\begin{equation}\label{eps-1}
\tilde{\epsilon}^{-1}_{{\bf q},{\bf k}}(\omega)\!=\!\delta_{{\bf
q},{\bf k}}
\!-\!\int \! d{\bf p}\,\chi_{{\bf q},{\bf p}}(\omega)\!
\left[(4 \pi / p^2)\,\delta_{{\bf p},{\bf k}}\!+\!
f^{xc}_{{\bf p},{\bf k}}(\omega)\!\right],
\end{equation}
and $f^{xc}_{{\bf q},{\bf k}}(\omega)$ is the Fourier transform of
the xc kernel
\begin{equation}\label{fxc}
f^{xc}[n_0]({\bf r},{\bf r}';\omega)=\left.{\delta v^{xc}[n]({\bf
r},\omega) / \delta n({\bf r}',\omega)}\right|_{n=n_0},
\end{equation}
with $v_{xc}[n]({\bf r},\omega)$ being the frequency-dependent xc
potential. $\chi^0_{{\bf q},{\bf k}}(\omega)$ is the Fourier
transform of the density-response function of {\it noninteracting}
KS electrons, i.e., independent electrons in the
effective KS potential
\begin{equation}\label{pot}
\tilde v({\bf r})= -Z_1/r+\int d{\bf r}'n_0({\bf r}')/|{\bf r}-{\bf
r}'|+v^{xc}[n_0]({\bf r}),
\end{equation}
with $v^{xc}[n_0]({\bf r})$ being the static xc potential at the
electron density $n_0({\bf r})$. The  unperturbed density
$n_0({\bf r})$  is
obtained by solving self-consistently the KS equations of DFT
with the effective potential $\tilde v({\bf r})$ of Eq.~(\ref{pot}).
At $\omega=0$, we find the following sum rules:
\begin{equation}
\label{sum}
\int d{\bf q}\,{\bf q}\left(\begin{array}{c}
\tilde v_{\bf q}\,\chi^0_{{\bf q},{\bf k}}(0) \\
v_{\bf q}\,\chi_{{\bf q},{\bf k}}(0) \\
v_{\bf q}\,\tilde{\epsilon}^{-1}_{{\bf q},{\bf k}}(0) \\
\end{array} \right)= {\bf k}
\left(
\begin{array}{r}
-n^0_{\bf k} \\
-n^0_{\bf k} \\
\tilde v_{\bf k} \\
\end{array}
\right),
\end{equation}
where $n^0_{\bf q}$ and $\tilde v_{\bf q}$ are the Fourier transforms
of the  density $n_0({\bf r})$ and the
potential $\tilde v({\bf r})$. With the aid of these sum
rules, substitution of Eq.~(\ref{eq}) into Eq.~(\ref{stopp}) yields
\begin{equation}\label{s}
Q = Q_1 + Q_2,
\end{equation}
where
\begin{eqnarray}
\label{res1}
Q_1 \!= \! \! \frac{1}{ 8\pi^3}
\int \! \! \! d{\bf q}  d{\bf k} \, \tilde v_{\bf q} \tilde v_{\bf k}
\frac{({\bf q}\cdot{\bf v})}{v} \! \! \frac{({\bf k}\cdot{\bf v})}{v}
\frac{\partial {\rm Im} \chi^0_{{\bf q},{\bf k}}(\omega)}{\partial\omega}  \Bigr |_{\omega=0},&&\\
\label{res2}
Q_2 \! =  \! \! - \! \frac{1}{8\pi^3} \! \!
\int \! \! \! d{\bf q}  d{\bf k} \, n^0_{\bf q}  n^0_{\bf k} \!
\frac{({\bf q}\cdot{\bf v})}{v} \! \! \frac{({\bf k}\cdot{\bf v})}{v}
\frac{\partial {\rm Im} f^{xc}_{{\bf q},{\bf k}}(\omega)}{\partial\omega} \!
\Bigr |_{\omega=0}. &&
\end{eqnarray}

Equations~(\ref{s})-(\ref{res2}), which  describe the SP of an
interacting many-electron system in the low-velocity limit, show
that the low-velocity SP of {\it interacting} electrons
[Eq.~(\ref{stopp})] can be obtained as the sum of (i) the SP of
{\it noninteracting} KS electrons [Eq.~(\ref{res1})] and
(ii) an excitonic contribution [Eq.~(\ref{res2})], which accounts
for the {\em dynamical} correlation that is absent in Eq.~(\ref{res1}).

In order to show that the SP of Eq.~(\ref{res1}) is precisely the SP
reported in Ref.~\onlinecite{Echenique-81} under the assumption of
individual elastic electron scattering by the effective KS
potential  of Eq.~(\ref{pot}), we write an explicit expression for
the imaginary part of $\chi^0_{{\bf q},{\bf k}}(\omega)$ by using
incoming and outgoing single-particle states, $| {\bf p'}^{(-)}\rangle$
and $| {\bf p}^{(+)}\rangle$, and the corresponding single-particle
energies, $\varepsilon_{\bf p'}$ and $\varepsilon_{\bf p}$
\footnote{Typically, the same set of eigenfunctions is used for both bra
and ket vectors entering Eq.~(\ref{imchi0}). Nevertheless, it can be shown
that two different orthonormal complete sets can be used, which we here
choose to be the incoming and outgoing states. Although we work with the
density-response function of an inhomogeneous EG rather than the
force-force correlation function, our proof has much in common with a
derivation in: E.~G.~d'Agliano, P.~Kumar, W.~Schaich, and H.~Suhl,
Phys. Rev. B {\bf 11}, 2122 (1975).}
\begin{eqnarray}
{\rm Im} \, \chi^0_{{\bf q},{\bf k}}(\omega)\!&=&\! (2\pi)^{-2}
\!\int \! d {\bf p}d {\bf p}'[f(\varepsilon_{{\bf p}'})\!-\!
f(\varepsilon_{\bf p})]\delta(\omega\!-\!\varepsilon_{\bf p}
\!+\!\varepsilon_{{\bf p}'})\cr\cr
&\times&
\langle {\bf p'}^{(-)}|\, e^{i {\bf q r}}\,| {\bf p}^{(+)}\rangle
\langle
{\bf p}^{(+)}|\,e^{-i {\bf k r}}\,|{\bf p'}^{(-)}\rangle,
\label{imchi0}
\end{eqnarray}
where $f(\varepsilon)$ denotes the Fermi distribution function.
Substituting Eq.~(\ref{imchi0}) into Eq.~(\ref{res1}) and referring to
the Lipp-man-Schwinger equation of scattering theory, one finds
\begin{equation}
Q_1
= \bar n  \, k_F \sigma_{tr}(k_F),
\label{start1}
\end{equation}
where $k_F$ is the Fermi momentum, $\sigma_{tr}(q)$ is
the so-called transport cross section, and $\bar n$ is the density of
the homogeneous EG in the absence of the projectile.

\begin{figure*}
\includegraphics[width=0.29\textwidth,height=0.27\textwidth]{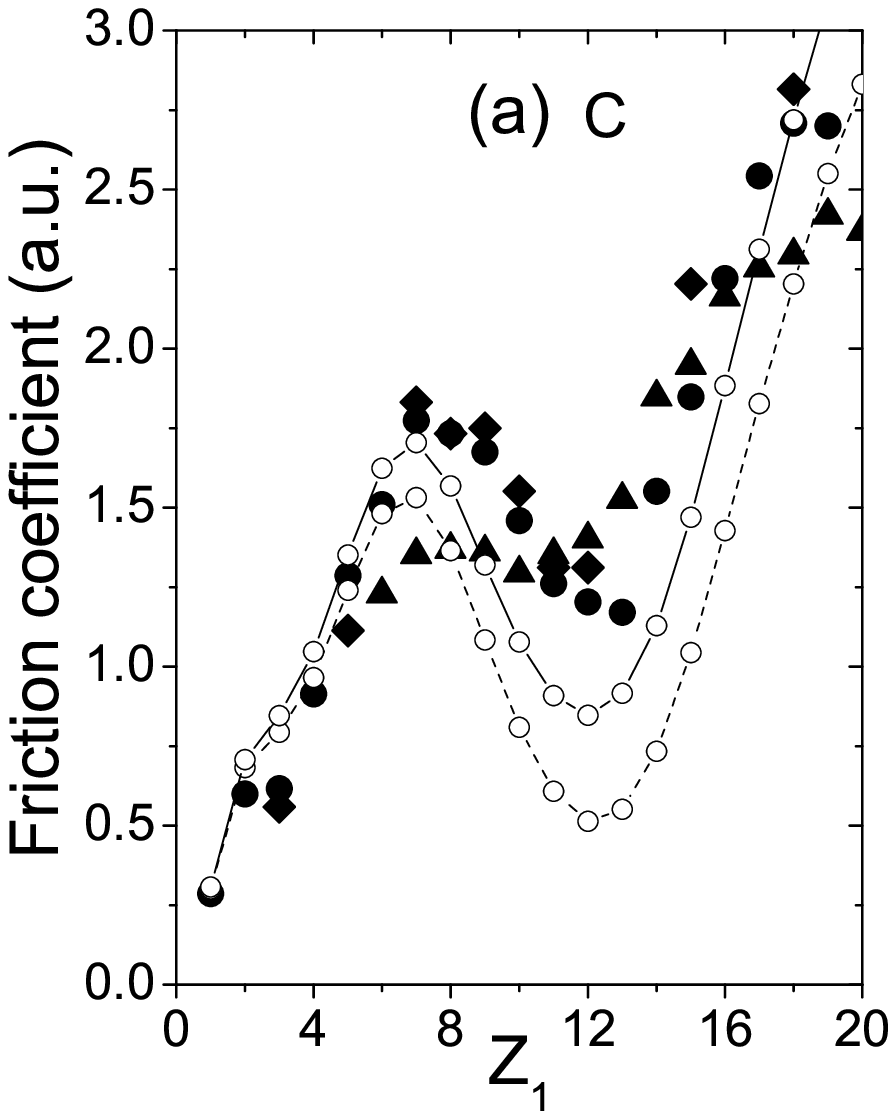}
\includegraphics[width=0.25\textwidth,height=0.27\textwidth]{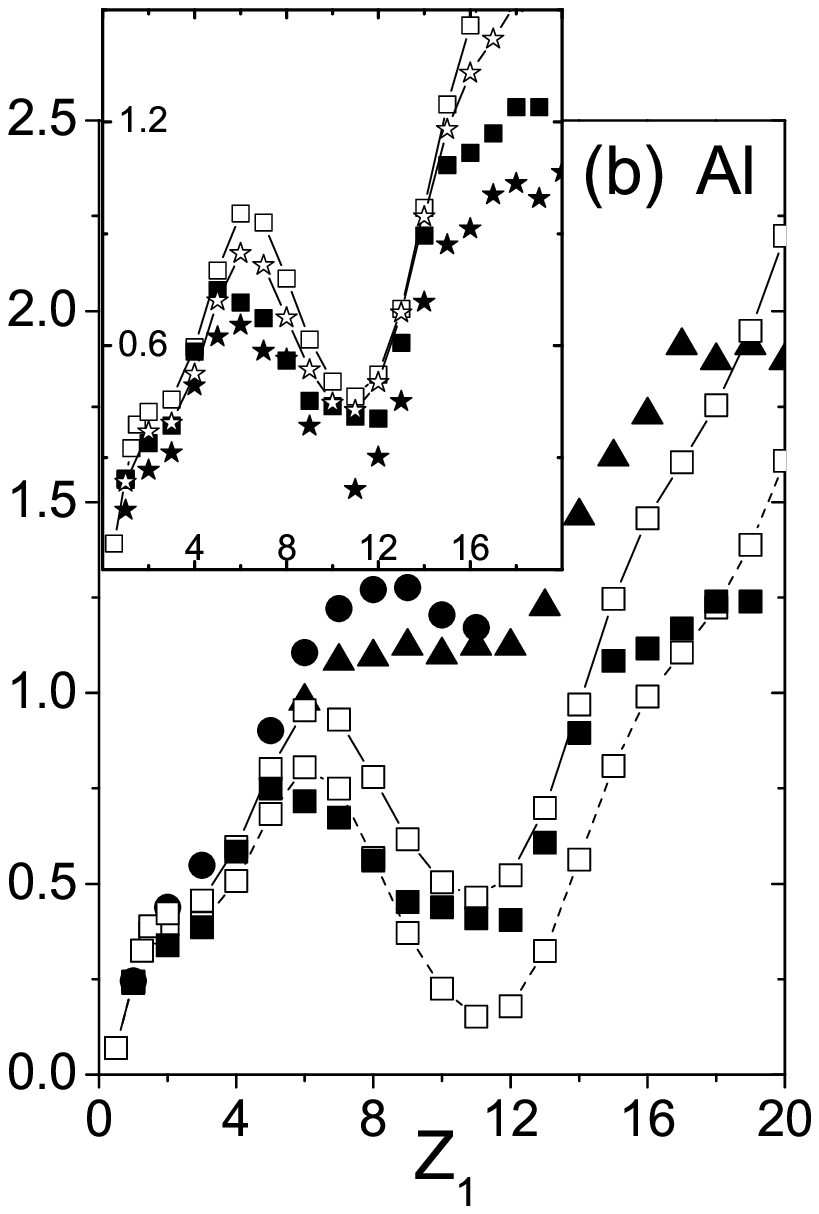}
\includegraphics[width=0.25\textwidth,height=0.27\textwidth]{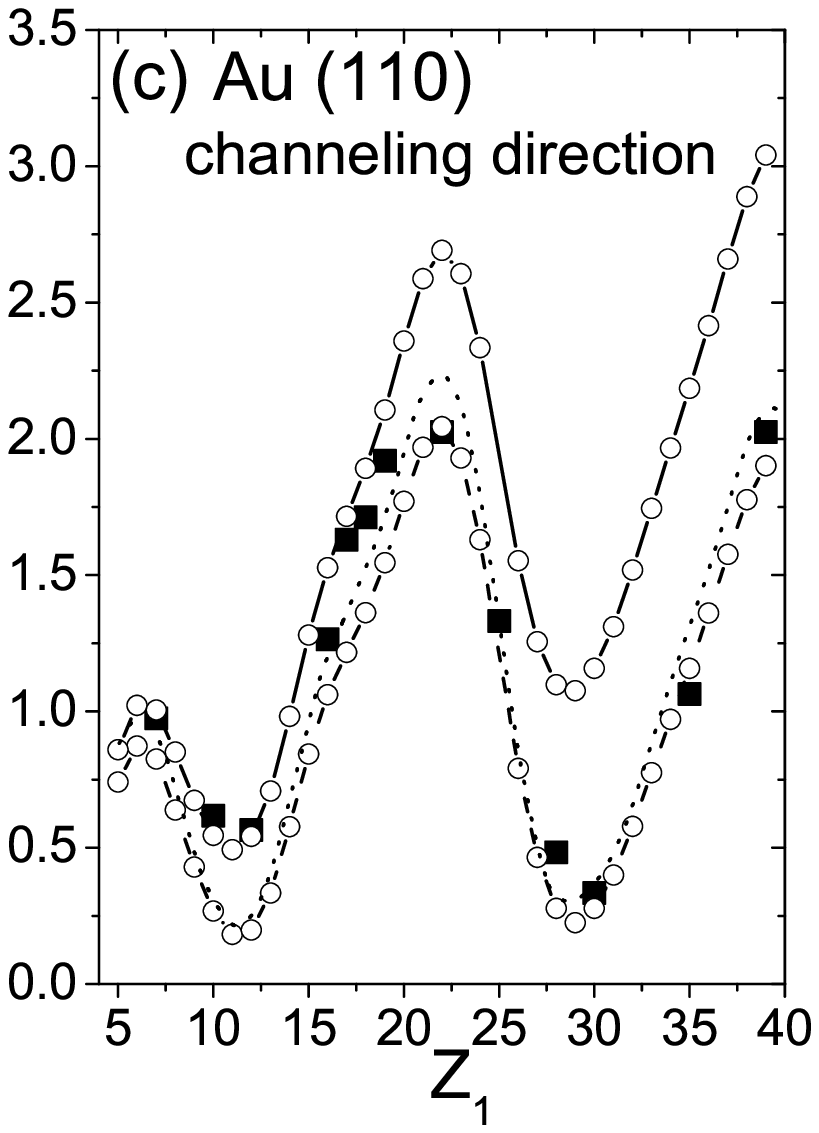}
\includegraphics[width=0.18\textwidth,height=0.27\textwidth]{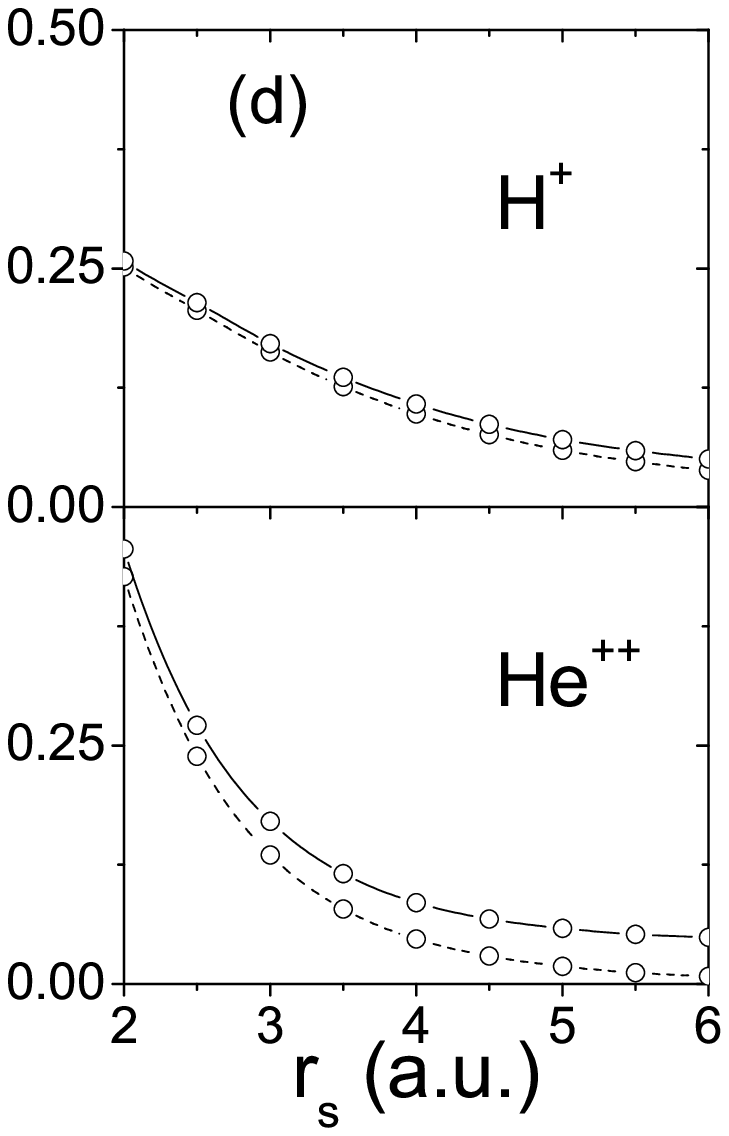}
\caption{Friction coefficient of a homogeneous EG versus the projectile
charge $Z_1$ (a)-(c) and the EG-density parameter $r_s$ (d), as
obtained in the LDA from Eq.~(\ref{start1}) (dashed chained curves)
and from the sum of Eqs.~(\ref{start1}) and (\ref{fxcl}) (solid
chained curves). (a) Solid circles, triangles, and diamonds are
transmission measurements from Refs.~\onlinecite{Ormrod-6365},
\onlinecite{Ward-79s}, and \onlinecite{Hogberg-71} of the random SP of C
for ions moving with velocities $v=0.41$, $0.83$, and $0.25
{\rm a.u.}$, respectively. (b) Solid circles and triangles are
transmission measurements from Refs.~\onlinecite{Ormrod-6365} and
\onlinecite{Ward-79s} of the random SP of Al for projectiles moving
with velocities $v=0.41$ and $0.83{\rm a.u.}$, respectively. Solid
squares (stars in the inset) are the measured SP of Al
of Ref.~\onlinecite{Winter-98s} (\onlinecite{Winter-03s}) for slow ions
($v=0.5{\rm a.u.}$) moving at a distance of 1.0 (1.2) a.u. from the last
atomic row at the surface. (c) Solid squares are the measurements
of Ref.~\onlinecite{Bottiger-69} of the SP of Au for slow ions
($v=0.68{\rm a.u.}$) channeled along the (110) direction. These
measurements are compared to EG calculations with $r_s=2$. The dotted line
is the calculation of Ref.~\onlinecite{Nagy2-89} with dynamical xc
effects included in the framework of linear-response theory.}
\label{figure1}
\end{figure*}

An evaluation of Eq.~(\ref{res1}) [or, equivalently, Eq.~(\ref{start1})]
and Eq.~(\ref{res2}) involves the {\it unknown} quantities
$v^{xc}[n_0]({\bf r})$ and $f^{xc}[n_0]({\bf r},{\bf r}';\omega)$,
respectively. Existing DFT calculations from Eq.~(\ref{start1}) have
been implemented by invoking the LDA, i.e., by replacing the xc potential
$v^{xc}[n_0]({\bf r})$ by the xc potential of a homogeneous EG with the
local density $n_0({\bf r})$. At the same level of approximation, we
use the LDA to evaluate the excitonic contribution of Eq.~(\ref{res2}).
In this approach, the frequency-dependent (nonadiabatic) xc kernel
$f^{xc}[n_0]({\bf r},{\bf r}';\omega)$ is obtained as\cite{Gross-85}
\begin{equation}\label{fxc0}
f^{xc}[n_0]({\bf r},{\bf r}';\omega)=f^{xc}_h[n_0({\bf r}),\omega]\,\delta({\bf
r}-{\bf r}'),
\end{equation}
where $f^{xc}_h(n,\omega)$ denotes the $q\to 0$ limit of the
frequency-dependent xc kernel of a homogeneous EG of
density $n$. Due to the spherical symmetry of $n_0({\bf r})$,
substitution of Eq.~(\ref{fxc0}) into Eq.~(\ref{res2}) yields
\begin{equation}
Q_2=
-\frac{4\pi}{3}  \int\limits_0^\infty
dr\left[r\,n_0'(r)\right]^2
\frac{\partial {\rm Im}f^{xc}_{h}[n_0(r),\omega]}
{\partial\omega}\Bigr |_{\omega=0}.
\label{fxcl}
\end{equation}

An explicit expression for the imaginary part of $f^{xc}_{h}(n,\omega)$
has been reported recently, which is exact at low frequencies to leading
order in the Coulomb interaction.\cite{Qian-02}
Since the frequency derivative entering Eq.~(\ref{fxcl}) is
negative definite,\cite{Qian-02} this
equation shows that dynamical correlation enhances the SP of a system
of interacting electrons. This enhancement can be ascribed to the
attractive e-h multiple scattering or excitonic effect in the e-h pair
excitation.\cite{Takada-02}

The calculations presented below have been carried out from
Eqs.~(\ref{start1}) and (\ref{fxcl}), with use of the LDA xc
potential $v^{xc}[n_0]({\bf r})$ of Perdew and
Zunger\cite{Perdew-81} and the frequency-dependent LDA xc kernel
$f^{xc}_{h}(n,\omega)$ of Qian and Vignale.\cite{Qian-02} We have
performed calculations for various values of $Z_1$ and the
electron-density parameter $r_s=(3/4\pi\bar n)^{1/3}$. We have
found that dynamical correlation can be neglected at small projectile
charges and high electron densities, where many-body effects play a
minor role. At metallic densities ($1.5<r_s<6$), however, the
correlation effect is found to considerably enhance the SP for
intermediately charged ions.

At low velocities, the energy loss of ions in solid materials is
mainly due to the stopping power of valence electrons. Since these
electrons are known to be well described by a homogeneous
EG,\cite{Campillo-98} we apply our theory to the cases of C, Al,
and Au, and compare the result of our EG calculations to the available
experimental data for these materials (see Fig. 1).

First, we compare the results we have obtained for the friction
coefficient of a homogeneous EG with $r_s=1.59$ and $r_s=2.07$
[solid chained curves in Figs. 1(a) and 1(b)] to the transmission
measurements of the random SP for slow ions in C and Al
(solid circles, squares and diamonds). If we neglect dynamical
correlation (dashed chained curves), our calculations reproduce
those of Ref.~\onlinecite{Echenique-86}. The effect of
dynamical correlation
is found to be negligible at low ions charges; however,
as $Z_1$ increases this effect becomes comparable to the SP of
noninteracting KS electrons. We observe
that the impact of dynamical correlation becomes relatively larger at
the formation of closed shells of bound states around the probe
particle, where the SP has a local minimum. As a result, the ratio
of the maximum to minimum values in the oscillations of the SP
decreases in the presence of dynamical correlation, leading
to better agreement with experiment.

Differences between our  calculation and the
transmission measurements of the random SP of C and Al should be
partially due to the contribution from elastic collisions of the
projectile with the atomic cores of the target, which is not
included in our calculation. Hence, we have also considered the
measured SP of Al for slow ions moving at a distance of $d=$1.0 a.u.
from the last atomic row at the surface\cite{Winter-98s} [solid squares
in Fig. 1(b)], which is free from elastic collisions with target atoms.
At this distance, the electron density is close to the bulk value
($r_s=2.07$) and the SP is expected to be only slightly smaller than
that for ions moving in a uniform EG.\cite{Garcialekue-01}
Figure 1(b) shows that this is indeed the case, although our
many-body calculation (solid chained curve) is too high for $Z_1>15$.
As $Z_1$ increases the electron-density variation near the projectile
becomes large, and differences between our calculations and the
experimental data may, therefore, be ascribed to the failure
of the LDA.

To elucidate the effect of the inhomogeneity of the EG near the Al
surface, in the inset of Fig.~1(b) we plot again our many-body
calculation at $r_s=2.07$ and the measured data at $d=$1.0 a.u., but
now together with the measured SP of Al  for slow ions moving at
$d=$1.2 a.u.\cite{Winter-03s} (solid stars) and the calculated curve
for the corresponding density of $r_s=2.2$ (solid chained curve with
stars). Since the two theoretical curves essentially coincide, this
figure shows that at 1.2 a.u., where the gradient of the electron
density is considerably larger than that at $d=$1.0
a.u.,\cite{Winter-03s} SP is worsly described by the  homogeneous
EG.

Contributions from elastic collisions with the target atoms should
also be absent in channeling experiments, in which the projectile
penetrates a solid along a high-symmetry direction. In Fig.~1(c)
we compare the experimental data of Ref.~\onlinecite{Bottiger-69}
(solid squares) for slow ions channeled along the (110) direction
in Au with the result of our many-body calculation for the
friction coefficient of a homogeneous EG with $r_s=2.0$. This figure
shows that many-body correlation effects tend to decrease the ratio
of the maximum to minimum values in the oscillatory behavior of the SP,
thus bringing the SP into nice agreement with experiment for projectile
charges up to $Z_1=$19.
However, comparison between theory and experiment at $Z_1>19$ suggests that
at these large values of $Z_1$ the strong inhomogeneity of the EG
suppresses dynamical correlations and calls for going beyond the LDA.

Finally, Fig. 1(d) displays the interacting and noninteracting SP
for H$^+$ and He$^{++}$, as a function of $r_s$. Figure 1(d) shows
that (i) the effect of dynamical correlation is very small in the case
of H$^+$, in which case the electron density variation is comparatively
small, and (ii) the dynamical correlation effect becomes comparatively
more important as the electron density decreases.

In conclusion, we have presented a rigorous TDDFT for the stopping power of
an {\it interacting} EG in the limit of low projectile velocities. In the
absence of dynamical correlation, our theory yields the SP of
{\it noninteracting} KS electrons, which we have shown to be
equivalent to the existing DFT-based elastic electron scattering approach.
We have investigated the effect of many-body dynamical correlation by
using an accurate representation of the long-wavelength frequency-dependent
xc kernel. Our results show that dynamical correlations considerably
enhance the SP of intermediately charged ions, thereby bringing the
theory into better agreement with the experimentally observed
oscillations of the SP for projectiles of charge $Z_1<$15. However,
at higher values of $Z_1$,  one needs to go beyond the LDA
in the description of the dynamical xc effects.

V.U.N. and C.S.K. acknowledge support by the Korea Research Foundation
through Grant No. KRF-2003-015-C00214. J.M.P. acknowledges partial
support by the UPV/EHU and the MCyT.

\end{document}